\documentstyle[12pt]{article}
\begin{document}
\begin{center}
{\Large \bf Recent Results in Charged-Composite Particle \\ Scattering}\\[4mm]
{\large \em Erwin O.\ Alt}\\[2mm]
{\small Institut f\"ur Physik, Universit\"at Mainz, D-55099 Mainz, Germany}\\
\end{center}
\normalsize

\begin{abstract}
A brief overview is given of some recent advances in charged-composite particle scattering. On the theoretical side, I address the three-charged particle wave function asymptotics, the nonperturbative investigation of the long-range behaviour of the optical potential, and the question of the compactness of the kernels of the momentum space integral equations for three charged particles. Among the more practical developments, I report on results of numerical calculations of so-called "triangle" amplitudes, a new, simple and very efficient higher-energy approximation for the latter, and a breakthrough in the quantitative treatment of Coulomb effects in proton-deuteron elastic scattering with realistic nuclear potentials. 
\end{abstract}

\section{Theoretical Developments}

\subsection{Asymptotic wave function for three free charged particles}

Knowledge of the asymptotic boundary condition on the three-free charged particle wave function is required not only when attempting to solve the Schr\"odinger equation above the ionization threshold, but also when investigating asymptotic properties of various interesting quantites like the optical potential or the behaviour of the kernel of momentum space integral equations. I, therefore, start by briefly recapitulating some important aspects of the asymptotic behaviour of the solutions of the Schr\"odinger equation.

Consider three distinguishable particles with masses 
$m_{\nu}$ and charges $e_{\nu},\,\nu = 1,2,3$. I use Jacobi coordinates:
${\rm {\bf {k}}}_{\alpha} \,({\rm {\bf {r}}}_{\alpha})$ are 
relative momentum (coordinate) between particles $\beta$ and $\gamma$, 
${\rm {\bf {q}}}_{\alpha} \,({\mbox {\boldmath $\rho$}}_{\alpha})$ the relative 
momentum (coordinate) between particle $\alpha$ and the center of mass 
of the pair $(\beta \gamma)$. The on-shell momenta are denoted by 
($\bar{\rm {\bf {k}}}_{\alpha}, \bar{\rm {\bf {q}}}_{\alpha}$) so that the on-shell relation reads as 
$E ={\bar q_{\alpha}}^{2}/2 M_{\alpha} + {\bar k_{\alpha}}^{2}/2 \mu_{\alpha}$, with 
$\mu_{\alpha} =m_{\beta}m_{\gamma}/(m_{\beta} +
m_{\gamma})$ and $M_{\alpha} = m_{\alpha}(m_{\beta}+m_{\gamma})/(m_{\alpha}
+m_{\beta}+m_{\gamma})$ being the appropriate reduced masses. Furthermore, 
$V_{\alpha} = V_{\alpha}^S + V_{\alpha}^C$ is the short-range plus Coulomb 
potential acting between particles $\beta$ and $\gamma$, and $\bar \eta_{\alpha} \equiv 
\eta_{\alpha}(\bar k_{\alpha}) = e_{\beta} e_{\gamma}\mu_{\alpha}/\bar k_{\alpha}$ the 
corresponding Coulomb parameter. Additional notation: 
$E_+ = E+i0$, $\epsilon_{\alpha \beta} = - \epsilon_{\beta \alpha } = +1$ if 
$(\alpha,\beta)$ = cyclic ordering of (1,2,3); finally, unit vectors are characterised by 
a hat: ${\hat{\rm{\bf{k}}}} = {\rm{\bf{k}}}/k$.

The asymptotic solutions of the Schr\"odinger equation for three asymptotically free particles 
in the various regions of configuration space are known: \\ 
In $\Omega _{0}$: $r_{1}, r_2, r_3 ~\rightarrow ~\infty$, but {\em not } 
$r_{\nu}/\rho_{\nu} \rightarrow 0 \mbox{ for }\, \nu = 1, 2, 3$ (Redmond, as cited in Ref.\ \cite{ros72}):
\begin{eqnarray} 
\Psi^{(+)}_{\bar {\rm {\bf {k}}}_{\alpha} \bar {\rm {\bf {q}}}_{\alpha}}
({\rm {\bf {r}}}_{\alpha} ,{\mbox {\boldmath $\rho$}}_{\alpha}) 
&\approx&
e^{ i (\bar {\rm {\bf {k}}}_{\alpha} \cdot {\rm {\bf {r}}}_{\alpha} + 
\bar {\rm {\bf {q}}}_{\alpha} \cdot {\mbox {\boldmath $\rho$}}_{\alpha})}
\prod_{\nu = 1}^3 e^{ i \bar \eta_{\nu} \ln(\bar k_{\nu} r_{\nu}-
\bar {\rm {\bf {k}}}_{\nu} \cdot {\rm {\bf {r}}}_{\nu})}. \label{psi0'}
\end{eqnarray} 
In $\Omega _{\alpha}$: $ \rho_{\alpha} \rightarrow \infty$, $r_{\alpha}/\rho_{\alpha} 
\rightarrow 0$, for $\alpha = 1,2$ or $3$ (\cite{am92}, 
with some refinements given in \cite{ks96,ml96}):
\begin{eqnarray}
\Psi^{(+)}_{\bar {\rm{\bf{k}}}_{\alpha} \bar {\rm{\bf{q}}}_{\alpha}}
({\rm {\bf {r}}}_{\alpha} ,{\mbox {\boldmath $\rho$}}_{\alpha}) \approx 
\psi^{(+)}_{\bar {\rm{\bf{k}}}_{\alpha}({\mbox {\boldmath $\rho$}}_{\alpha})}
({\rm {\bf{r}}}_{\alpha})\,
e^{ i \bar {\rm {\bf {q}}}_{\alpha} \cdot {\mbox {\boldmath $\rho$}}_{\alpha}}
\, \prod_{\nu \neq \alpha}e^{i \bar \eta_{\nu } \ln(\bar k_{\nu } \rho_{\alpha}-
\epsilon_{\alpha \nu } \bar {\rm{\bf{k}}}_{\nu } \cdot
{\mbox {\boldmath $\rho$}}_{\alpha})}.\label{psias}
\end{eqnarray}
Here, $\psi^{(+)}_{\bar {\rm{\bf{k}}}_{\alpha}({\mbox {\boldmath $\rho$}}_{\alpha})}
({\rm {\bf{r}}}_{\alpha})$ is continuum solution of the two-body-like Schr\"odinger equation 
\begin{eqnarray} 
\left\{ \frac{{\bar k_{\alpha}}^{2}({\mbox {\boldmath $\rho$}}_{\alpha})}
{2 \mu_{\alpha}} + \frac{\Delta_{{\rm {\bf{r}}}_{\alpha}}}{2 \mu_{\alpha}} -
V_{\alpha}({\rm {\bf{r}}}_{\alpha})\right\} \psi^{(+)}_{\bar {\rm{\bf{k}}}_{\alpha}
({\mbox{\boldmath $\rho$}}_{\alpha})}({\rm {\bf{r}}}_{\alpha}) = 0, \label{psis} 
\end{eqnarray}
describing the relative motion of particles $\beta$ and $\gamma$ 
with {\em local} energy $E_{\alpha}({\mbox {\boldmath $\rho$}}_{\alpha}) 
= {\bar k_{\alpha}}^{2}({\mbox {\boldmath $\rho$}}_{\alpha})/2 
\mu_{\alpha}$, where (e.g.\ $ \lambda_{\beta} = \mu_{\alpha}/m_{\gamma}$, 
with $\beta, \gamma \neq \alpha$)
\begin{eqnarray} 
\bar {\rm{\bf{k}}}_{\alpha}({\mbox {\boldmath $\rho$}}_{\alpha}) 
= \bar {\rm {\bf {k}}}_{\alpha} + 
\frac{{\rm{\bf{a}}}_{\alpha}({\hat {\mbox {\boldmath 
$\rho$}}}_{\alpha})}{\rho_{\alpha}}, \quad 
{\rm{\bf{a}}}_{\alpha}({\hat {\mbox {\boldmath $\rho$}}}_{\alpha}) 
= - \sum_{\nu \neq \alpha} \bar \eta_{\nu} \lambda_{\nu} \frac{\epsilon_{\alpha \nu}
{\hat {\mbox {\boldmath $\rho$}}}_{\alpha} - 
{\hat {\bar {\rm{\bf{k}}}}}_{\nu}}{1 - \epsilon_{\alpha \nu}
{\hat {\mbox {\boldmath $\rho$}}}_{\alpha} \cdot 
{\hat {\bar {\rm{\bf{k}}}}}_{\nu}}. \label{aa}
\end{eqnarray}
Its parametric dependence on ${\mbox {\boldmath $\rho$}}_{\alpha}$ is a 
manifestation of long-ranged three-body correlations; thus it is, in fact, a three-body
wave function, the influence of the third particle $\alpha$ however being confined to 
a shift of the relative momentum of particles $\beta$ and $\gamma$ from its asymptotic 
(for $\rho_{\alpha} \to \infty$, i.e. particle $\alpha$ is infinitely far apart) value 
$\bar {\rm{\bf{k}}}_{\alpha}$ to the {\em local} value 
$\bar {\rm{\bf{k}}}_{\alpha}({\mbox {\boldmath $\rho$}}_{\alpha})$.

\subsection{Long-Range Behaviour of the Optical Potential in a Three-Body System}

Elastic scattering processes with (charged) composite particles of the type 
$\alpha + (\beta,\gamma)_m \to \alpha + (\beta,\gamma)_m$ can formally be described 
by means of a single one-channel LS equation for the elastic scattering amplitude
\begin{eqnarray}
{\cal T}_{\alpha m, \alpha m}(z) = {\cal V}_{\alpha m,
\alpha m}^{opt}(z) + {\cal V}_{\alpha m, \alpha m}^{opt}(z)
\frac{1}{z - {\rm {\bf {Q}}}_{\alpha}^{\,2}/2 M_{\alpha}
+ \hat E_{\alpha m}}{\cal T}_{\alpha m, \alpha m}(z),
\label{ie}
\end{eqnarray}
where the plane wave matrix elements of the optical potential (OP) operator are given as
\begin{eqnarray}
{\cal V}_{\alpha m, \alpha m}^{opt}({\rm {\bf {q}}}_{\alpha}',{\rm {\bf {q}}}_{\alpha};z) 
&=& \langle{\rm {\bf {q}}}_{\alpha}'|\langle \psi_{\alpha m} | 
\bar V_{\alpha} + \bar V_{\alpha} Q_{\alpha m}G(z) Q_{\alpha m} 
\bar V_{\alpha} |\psi_{\alpha m} \rangle 
|{\rm {\bf {q}}}_{\alpha} \rangle \nonumber \\[3mm]
&=& \underbrace{{\cal V}_{\alpha m, \alpha m}^{stat}({\rm{\bf{q}}}_{\alpha}', 
{\rm{\bf{q}}}_{\alpha})}_{\mbox{static potential}} + 
\underbrace{{\tilde {\cal V}}_{\alpha m, \alpha m}^{opt}
({\rm{\bf{q}}}_{\alpha}', {\rm{\bf{q}}}_{\alpha}; z)}_{\mbox{nonstatic part of OP}}.
 \label{vopt}
\end{eqnarray}
The following notation is used: $\mid \!\! \psi_{\alpha m} \rangle$ denotes the target bound 
state wave function to binding energy $\hat E_{\alpha m}$; $G(z) = (z-H_0 - \sum_{\nu} 
V_{\nu})^{-1}$ and $G^C(z)= (z-H_0 - \sum_{\nu} V_{\nu}^C)^{-1}$ are the resolvents 
of the full and the pure Coulomb three-body Hamiltonian,  and $G_{\alpha}(z) = 
(z-H_0 - V_{\alpha})^{-1}$ the one of the channel Hamiltonian. 
$P_{\alpha m} = |\psi_{\alpha m}\rangle \langle \psi_{\alpha m}|$ 
projects onto the target state, and $Q_{\alpha m}= 1 - P_{\alpha m}$ onto the orthogonal 
complement. Furthermore, $\bar V_{\alpha} = \bar V_{\alpha}^S+\bar V_{\alpha}^C = 
\sum_{\nu \neq \alpha} V_{\nu}$ is the channel interaction; ${\rm {\bf {Q}}}_{\alpha}$
is the momentum operator with eigenvalue ${\rm {\bf {q}}}_{\alpha}$.

The question of solvability of (\ref{ie}) depends on the singular behaviour 
of ${\cal V}_{\alpha m, \alpha m}^{opt}$ in the limit that the momentum 
transfer ${\rm {\bf {\Delta}}}_{\alpha} = {\rm {\bf {q}}}_{\alpha}'-
{\rm {\bf {q}}}_{\alpha}$ goes to zero. In leading order, the latter is  
caused by the Coulombic part $\bar V_{\alpha}^C$ of the 
channel interaction, i.e., it does not depend on either the 
short-range part $\bar V_{\alpha}^S$ or on the internal 
interaction $V_{\alpha}$. Of course, a certain behaviour of the 
optical potential for $\Delta_{\alpha} \rightarrow 0$ implies a 
corresponding asymptotic behaviour in coordinate space.

As is well known, the static potential has, in the limit $\Delta_{\alpha} \to 0$, only 
the trivial Coulomb-type singular behaviour, which for a spherically symmetric target looks like: 
\begin{eqnarray} 
{\cal V}_{\alpha m, \alpha m}^{stat} ({\rm{\bf{q}}}_{\alpha}',{\rm{\bf{q}}}_{\alpha}) 
\stackrel{ \bar\Delta_{\alpha}\rightarrow 0}{=} 
\frac{4\pi e_{\alpha}(e_{\beta}+e_{\gamma})}{\Delta_{\alpha}^{2}} \quad \Longleftrightarrow \quad 
{\cal V}_{\alpha m, \alpha m}^{stat}({\mbox{\boldmath $\rho$}}_{\alpha})
\stackrel{\rho_{\alpha} \rightarrow \infty}{=} 
\frac{e_{\alpha}(e_{\beta}+e_{\gamma})}{ \rho_{\alpha}}. \label{as0}
\end{eqnarray}

The behaviour of the nonstatic OP has been known for a long time, but only in $2^{nd}$ 
order perturbation theory which in the present language is equivalent to approximating in 
(\ref{vopt}) the three-body ($G$) by the channel resolvent ($G_{\alpha}$), or in adiabatic 
approximation. On the energy 
shell and below the ionisation threshold one has (with ${\bar {\rm {\bf {\Delta}}}}_{\alpha}
= \bar {\rm{\bf{q}}}_{\alpha}' - \bar {\rm {\bf {q}}}_{\alpha}$)
\begin{eqnarray} 
{\tilde {\cal V}}_{\alpha m, \alpha m}^{opt\,(2)}(\bar{\rm{\bf{q}}}_{\alpha}', 
\bar{\rm{\bf{q}}}_{\alpha}; E_+) \stackrel{ \bar\Delta_{\alpha}\rightarrow 0}{\sim} 
\bar\Delta_{\alpha} \qquad \Longleftrightarrow \qquad {\tilde {\cal V}}_{\alpha m, \alpha m}^{opt\,(2)}
({\mbox{\boldmath $\rho$}}_{\alpha})\stackrel{\rho_{\alpha} \rightarrow \infty}
{\approx}- \frac{a}{2 \rho_{\alpha}^4}. \label{as0'}
\end{eqnarray}
Here, $a$ is the so-called static dipole polarisability of the composite particle. Two questions arise 
immediately: \\
(i) Does the fundamental result (\ref{as0'}) hold also for the exact nonstatic OP, 
i.e., even after all terms of the perturbation expansion of 
$G(z)$ are summed up? (For $E < 0$, this has been answered in the affirmative, though 
not fully rigorously, in \cite{km88}.) \\
(ii) Does it hold also for $E > 0$? (Within perturbative approaches the answer was 
yes, but with an energy-dependent $"a"$ in \cite{mss82}, and with the standard $a$ in \cite{uc89}.)

The behaviour of ${\tilde {\cal V}}_{\alpha m, \alpha m}^{opt}$ for $ \Delta_{\alpha}\rightarrow 0$ 
has been investigated in \cite{am95} nonperturbatively and for all energies, by inserting the 
spectral decomposition of $ G^C(E_+)$ with both two- and three-body intermediate states; 
for the latter the asymptotic three-charged particle wave function in $\Omega_{\alpha}$ has 
been used. Off the energy shell we find:
\begin{eqnarray} 
{\tilde {\cal V}}_{\alpha m, \alpha m}^{opt}
({\rm{\bf{q}}}_{\alpha}', {\rm{\bf{q}}}_{\alpha}; E) 
\stackrel{{\Delta}_{\alpha} \rightarrow 0}{=}
C_{1} \Delta_{\alpha} + o\,(\Delta_{\alpha}), \label{vas5}
\end{eqnarray}
with
\begin{eqnarray} 
C_{1} = -\frac{\pi^2}{4}\left\{\sum_{n \neq m}
\frac{\left[|{\rm{\bf{D}}}_{nm}|^2 + |\hat {\rm {\bf {\Delta}}}_{\alpha}
\cdot {\rm{\bf{D}}}_{nm}|^2 \right]}{[E_+ - q_{\alpha}^2/2 M_{\alpha} - 
\hat E_{\alpha n}]} + \int \frac{d {\rm{\bf{k}}}_{\alpha}^{0}}{(2\pi)^{3}}
\frac{\left[|{\rm{\bf{D}}}_{{\rm{\bf{k}}}_{\alpha}^{0} m}|^2 + 
|\hat {\rm {\bf {\Delta}}}_{\alpha} \cdot 
{\rm{\bf{D}}}_{{\rm{\bf{k}}}_{\alpha}^{0} m}|^2 \right]}
{[E_+ -q_{\alpha}^2/2 M_{\alpha} -
{k_{\alpha}^{0}}^2/2 \mu_{\alpha}]}\right\}, \label{cmm}
\end{eqnarray}
and ($N_{\alpha} = \epsilon_{\alpha \beta} e_{\alpha}\mu_{\alpha} 
[e_{\gamma}/m_{\gamma} - e_{\beta}/m_{\beta}]$)
\begin{eqnarray} 
{\rm{\bf{D}}}_{nm}= N_{\alpha} \int d {\rm{\bf{r}}}_{\alpha} 
\psi_{\alpha n}^{*}({\rm{\bf{r}}}_{\alpha}) {\rm{\bf{r}}}_{\alpha} \psi_{\alpha m}
({\rm{\bf{r}}}_{\alpha}), \quad {\rm{\bf{D}}}_{{\rm{\bf{k}}}_{\alpha}^{0} m}
=N_{\alpha}\int d {\rm{\bf{r}}}_{\alpha} \psi_{{\rm{\bf{k}}}_{\alpha}^{0}}^{(+)*}
({\rm{\bf{r}}}_{\alpha}) {\rm{\bf{r}}}_{\alpha} \psi_{\alpha m}({\rm{\bf{r}}}_{\alpha}).
\label{dkm2}
\end{eqnarray} 
On the energy shell, ${\tilde {\cal V}}_{\alpha m, \alpha m}^{opt}$ is seen to depend on the 
momenta only via ${\bar {\rm {\bf {\Delta}}}}_{\alpha}$, and no longer on the energy. Hence, 
in coordinate space it is a local, energy-independent potential
\begin{eqnarray} 
 {\tilde {\cal V}}_{\alpha m, \alpha m}^{opt}({\mbox {\boldmath $\rho$}}_{\alpha}) 
\stackrel{\rho_{\alpha} \to \infty}{=} - \frac{a}{2 \rho_{\alpha}^4} 
+ o\left( \frac{1}{\rho_{\alpha}^4}\right) , \label{coo}
\end{eqnarray}
with the polarisability as known from the perturbative approaches:
\begin{eqnarray} 
a=2 \sum_{n \neq m}\frac{|\hat{\mbox{\boldmath $\rho$}}_{\alpha} 
\cdot {\rm{\bf{D}}}_{nm}|^2}{\left[ |\hat E_{\alpha m}| - |\hat E_{\alpha n}| \right]} 
+2 \int \frac{d {\rm{\bf{k}}}_{\alpha}^0}
{(2\pi)^{3}}\frac{|\hat{\mbox{\boldmath $\rho$}}_{\alpha} 
\cdot {\rm{\bf{D}}}_{{\rm{\bf{k}}}_{\alpha}^0 m}|^2 }
{\left[ |\hat E_{\alpha m}| + {k_{\alpha}^0}^2/2 \mu_{\alpha} \right]}.
\label{adip}
\end{eqnarray}
That is, no "renormalisation" of $a$ arises from the higher order 
terms in the perturbation expansion of $G(z)$, and all dependence on $E$ 
which is present in the off-shell "strength factor" $C_{1}(E)$ has 
disappeared. Note that along the same lines the existence of a new nonrelativistic 
contribution to ${\tilde {\cal V}}_{\alpha m, \alpha m}^{opt}$ $\sim 1/\rho^5$ has 
been established recently in \cite{m97}.

\subsection{Compactness properties of the kernels of momentum space integral equations 
for three charged particles}

Three-body integral equations of the Faddeev type can not 
be used for Coulomb-like potentials, because of the occurrence of 
singularities in their kernels which destroy the compactness properties 
known to exist for short-range interactions. Up to now, only for energies 
below the breakup threshold had attempts been successful to obtain 
integral equations with compact kernels, by singling out from the original kernel 
the so-called two-particle Coulomb singularity, in a form such that it could be 
inverted explicitly \cite{ves70}. 

To investigate the behaviour of the kernels for positive energies, 
we use the rigorously equivalent formulation in terms of an effective-two-body 
theory \cite{ags67}. Here, the transition amplitudes for all binary processes 
satisfy a closed set of coupled LS-type equations 
\begin{eqnarray}
{\cal T}_{\beta n, \alpha m}(z) = {\cal V}_{\beta n, \alpha m}(z) + \sum_{\nu=1}^3\; 
\sum_{r,s}{\cal V}_{\beta n, \nu r}(z) {\cal G}_{0;\nu, rs}(z)\, {\cal T}_{\nu s, \alpha m} (z).
\label{ags1}
\end{eqnarray}
Without loss of generality, the short-range potentials can be 
taken as separable po\-ten\-tials: $V_{\alpha}^S = \sum_m\mid \!\! \chi_{\alpha m} \rangle 
\lambda_{\alpha m} \langle \chi_{\alpha m} \!\!\mid  $. 
The effective potential matrix elements are given as
\begin{eqnarray}
{\cal V}_{\beta n, \alpha m}({\rm {\bf{q}}}_{\beta}', 
{\rm {\bf{q}}}_{\alpha};z) = \langle {\rm {\bf{q}}}_{\beta}', 
\chi_{\beta n} \!\!\mid  G^{C}(z) - \delta_{\beta \alpha}G_{\alpha}^{C}(z) \mid \!\! 
\chi_{\alpha m},{\rm {\bf{q}}}_{\alpha} \rangle. \label{1efpot} 
\end{eqnarray}
The effective propagator matrix has simple poles in its 
diagonal for those $m$ which correspond to bound states: 
${\cal G}_{0;\alpha, mm} ({\rm {\bf{q}}}_{\alpha};z) \sim 
1/(z - q_{\alpha}^{2}/2M_{\alpha} - {\hat E}_{\alpha m})$. 
Note that as a result of our choice of the form of $V_{\alpha}^S$, 
these expressions contain only pure Coulombic quantities. 

Define $D_0(z) = z - {k}_{\beta}'^{2}/2\mu_{\beta}- {q}_{\beta}'^{2}/2M_{\beta} = 
z - {k}_{\alpha}^{2}/2\mu_{\alpha}- {q}_{\alpha}^{2}/2M_{\alpha}$, 
with $k_{\beta}' = |{\rm {\bf {q}}}_{\alpha} + \mu_{\beta}
{\rm {\bf {q}}}_{\beta}'/m_{\gamma}|, \, k_{\alpha} = |{\rm {\bf {q}}}_{\beta}' + \mu_{\alpha}
{\rm {\bf {q}}}_{\alpha}/m_{\gamma}|$, $\eta_{\beta} \equiv \eta_{\beta}(k_{\beta}')$ 
and $\eta_{\alpha} \equiv \eta_{\alpha}(k_{\alpha})$. Then, we have now proved 
\cite{maa97} that even for positive energies, provided all three particles have charges of 
equal sign, the leading singularity of the nondiagonal off-shell effective potential 
is a branch point at $D_0(z)= 0$: 
\begin{eqnarray}
{\cal V}_{\beta n, \alpha m}({\rm {\bf{q}}}_{\beta}', {\rm {\bf{q}}}_{\alpha};z) 
\stackrel{D_0(z) \to 0}{\sim} 1/D_0(z)^{(1 + i \eta_{\beta} + i \eta_{\alpha})}, 
\mbox{  for  } \beta \neq \alpha. \label{J2s2}
\end{eqnarray}
This compares with the simple pole occurring in the `pole' amplitude 
${\cal V}_{\beta n, \alpha m}^{pole}({\rm {\bf{q}}}_{\beta}', 
{\rm {\bf{q}}}_{\alpha};z)  = \bar \delta_{\beta \alpha} \langle {\rm {\bf{q}}}_{\beta}', 
\chi_{\beta n} \!\!\mid G_{0}(z) \mid \!\! \chi_{\alpha m},{\rm {\bf{q}}}_{\alpha} \rangle
\sim 1/D_0(z)$. On the energy shell, it goes over into
\begin{eqnarray}
{\cal V}_{\beta n, \alpha m}(\bar {\rm {\bf q}}_{\beta}', \bar {\rm {\bf q}}_{\alpha};E_+) 
\stackrel{\bar D_0 \to 0}{\sim} 1/{\bar D_0}^{(1- \eta^{(bs)}_{\alpha m}-\eta^{(bs)}_{\beta n})}, 
\mbox{  for  } \beta \neq \alpha, \label{tildens1}
\end{eqnarray}
where we have defined the bound state Coulomb parameter via 
$\eta_{\alpha m}^{(bs)} = e_{\beta}e_{\gamma}\mu_{\alpha}/\kappa_{\alpha m}$, 
with $\kappa_{\alpha m}^2 = {2\mu_{\alpha} |{\hat E}_{\alpha m}|}$, and similarly for 
$\eta_{\beta n}^{(bs)}$, and denoted by 
$\bar D_0 = {{\bar k}_{\alpha}}^{2}/2 \mu_{\alpha} + 
|\hat E_{\alpha m}|$ the on-shell restriction of $D_0(z)$. 
Since neither the pole nor the branch point singularity can coincide, 
for real values of the momenta, with the effective propagator pole, both are harmless.

Preliminary results indicate that the diagonal kernels develop on the energy shell 
a nonintegrable singularity which is, however, of same two-body type 
("center-of-mass Coulomb singularity") as that found in \cite{ves70} for $E<0$, 
and in \cite{asz78,as80} (see also \cite{as96}) for all energies within the screening 
approach; thus, it can be treated by explicit inversion. Apart from that the next strongest 
singularity occurs off shell and turns out to be $O(\Delta_{\alpha}^{-5/2})$ for 
$\Delta_{\alpha} \to 0$. Thus, it is integrable. 

Hence, it appears that after a few iterations the kernels are compact even above the 
breakup threshold provided the charges of all particles are of equal sign. In contrast, 
if some of the charges have opposite signs the kernels 
develop nonintegrable singularities which can destroy the compactness properties.

\section{Practical Developments}

\subsection{Energetic Collisions of Charged Projectiles with Atomic Bound States} 

Scattering of an elementary charged projectile off a two-charged particle 
bound state such as hydrogen atoms, positronium, etc.,
at higher energies can be described by the first few terms 
of the multiple-scattering expansion of the three-body transition operator:
\begin{eqnarray}
{\cal T}_{\beta n, \alpha m}(\bar {\rm {\bf {q}}}_{\beta}', \bar {\rm {\bf {q}}}_{\alpha}) 
= \langle \bar {\rm {\bf {q}}}_{\beta}' , \psi_{\beta n} \!\!\mid  \{\bar \delta_{\beta \alpha} G_0(E_+) 
+ \sum_{\nu}\bar \delta_{\beta \nu}\bar \delta_{\nu\alpha} T_{\nu}^C(E_+) + \cdots \}
\mid \!\! \psi_{\alpha m} , \bar {\rm {\bf {q}}}_{\alpha} \rangle . \label{ton}
\end{eqnarray}
Here, $G_0(E_+) = (E+ i0- H_0)^{-1}$, $T_{\gamma}^C$ the two-body 
Coulomb T-operator acting between particles $\alpha$ and $\beta$, and 
$\bar\delta_{\beta \alpha} = 1 - \delta_{\beta \alpha}$. The momenta satisfy 
the on-shell condition $E ={{\bar q}_{\alpha}}^{2}/2 M_{\alpha} 
+ \hat E_{\alpha m} = {{\bar q}_{\beta}}'^{2}/2 M_{\beta} + \hat E_{\beta n}$. 

The first (`pole') term of (\ref{ton}) describes the elementary one-particle transfer.
The first-order rescattering (`triangle') amplitudes ($\vartheta = 
\angle(\bar {\rm{\bf{q}}}_{\beta}', \bar {\rm{\bf{q}}}_{\alpha}))$ 
\begin{eqnarray}
\sum_{\nu}\bar \delta_{\beta \nu}
\bar \delta_{\nu\alpha} \langle \bar {\rm {\bf {q}}}_{\beta}' \mid 
\langle \psi_{\beta n} \mid T_{\nu}^C(E+i0)\mid \psi_{\alpha m} \rangle \mid 
\bar {\rm {\bf {q}}}_{\alpha} \rangle = 
\left \lbrace \begin{array}{l}
{\cal M}_{\beta n, \alpha m}^{T^C}(\vartheta, E) \quad \mbox{if} 
\quad \beta \neq \alpha \\
\sum_{\nu \neq \alpha}{\cal M}^{T^C}_{\nu, n m}
(\vartheta, E) \quad \mbox{if} \quad \beta = \alpha, 
\end{array}\right. 
\end{eqnarray}
contribute to both direct ($\beta = \alpha$) and exchange ($\beta \neq \alpha$) scattering. 
They (as well as all higher-order contributions) contain $T^C$ describing the rescattering 
of projectile $\alpha$ off each charged target particle. Hence, their calculation is difficult and 
time-consuming (this is particularly so if the rescattering particles have charges of opposite sign, 
because of the occurrence of the infinity of bound states). Thus, usually the approximation 
$T_{\nu}^C \to V_{\nu}^C$ is made, yielding the so-called Coulomb-Born 
Approximation (CBA), to be denoted by ${\cal M}_{\beta n, \alpha m}^{V^C}$ 
and ${\cal M}^{V^C}_{\nu, n m}$. But for atomic reactions the CBA is known 
to fail badly (for attractive and repulsive cases), 
except for $E \to \infty$ (see \cite{akm96} and references therein).

We have succeeded to numerically calculate all (direct and exchange) triangle 
amplitudes for all $E$ and $\vartheta$ for arbitrary wave functions, 
for repulsive and attractive intermediate-state rescattering (for the latter, a 
`new' representation of the attractive Coulomb T-matrix for 
$E<0$ was developed which has the bound state poles displayed explicitly in a simple, 
numerically convenient manner). Many concrete results are discussed 
in \cite{akm96,akmr95}. 

Moreover, we have derived a new approximation in the form of a `renormalized' CBA:
\begin{eqnarray} 
{\cal M}^{T^C}_{\gamma, n m}(\vartheta, E) 
\approx {\cal R}_{\gamma,n m}^{(s)}(\vartheta, E) \,{\cal M}^{V^C}_{\gamma, n m}
(\vartheta, E) , \quad \gamma \neq \alpha, \quad \\ \label{tappr}
{\cal M}^{T^C}_{\beta n, \alpha m}(\vartheta, E) 
\approx  {\cal R}_{\beta n, \alpha m}^{(s)}(\vartheta, E) \,{\cal M}^{V^C}_{\beta n, \alpha m}
(\vartheta, E), \quad \beta \neq \alpha,  \label{mappr3}
\end{eqnarray}
with 
\begin{eqnarray}
{\cal R}_{\gamma, n m}^{(s)}(\vartheta,E)=\frac{A^{i\eta_{\gamma}^{(d)}}
{\tilde {\cal R}}_{\gamma,n m}^{(s)}(E)}
{\Delta_{\alpha \alpha}^{2i \eta_{\gamma}^{(d)}} \left[\lambda_{\alpha}^2
\Delta_{\alpha \alpha}^2+ (\kappa_{\alpha n}+\kappa_{\alpha m})^2 \right]^
{- 2 i \eta_{\gamma}^{(d)}}}, \quad \gamma \neq \alpha, \\
{\cal R}_{\beta n, \alpha m}^{(s)}(\vartheta,E)=\frac{B^{i\eta_{\gamma}^{(e)}} 
{\tilde {\cal R}}_{\beta n, \alpha m}^{(s)}(E)}
{\left[\lambda_{\beta}^2\Delta_{\beta \alpha}^2+ (\kappa_{\beta n}+
\kappa_{\alpha m})^2 \right]^{- 2 i \eta_{\gamma}^{(e)}}}, \quad \gamma \neq \beta \neq \alpha.
\end{eqnarray}
Here, the following notation is used: 
${\rm{\bf{\Delta}}}_{\beta \alpha} = {\rm{\bf{q}}}_{\beta}'- \lambda_{\alpha}
{\rm{\bf{q}}}_{\alpha}/\lambda_{\beta}$, $\lambda_{\alpha}=\mu_{\alpha}/m_{\beta}, \, \lambda_{\beta}=\mu_{\beta}/m_{\alpha}$; $\kappa_{\alpha m}^2 = 2 \mu_{\alpha} 
|\hat E_{\alpha m}| $ and similarly for $\kappa_{\beta n}^2$. Finally, 
$\eta_{\gamma}^{(d,e)} = e_{\alpha}e_{\beta}\mu_{\gamma}/
\{2 \mu_{\gamma}(E_+ -k_{(d,e)}^2/2 M_{\gamma})\}^{1/2}$. The quantities 
$k_{(d)}(E), k_{(e)}(E)$ are real, $A(E),B(E)$ are real for $\eta_{\gamma}^{(d,e)}$ 
real, and ${\tilde {\cal R}(E)}$ is complex. All of them are independent of 
$\vartheta$, and are explicitly given in terms of simple functions. We note in parentheses 
that we have also derived approximate analytical expressions for ${\cal M}^{V^C}$ 
(for arbitrary bound state wave functions). When inserted into (\ref{mappr3}) they yield 
approximations for the triangle amplitudes containing no quadratures at all 
(however, their range of validity is somewhat limited). Because of their simple structure, 
both types are very easy to use for theoretical as well as numerical purposes. 

Selected applications: Consider $E > k_{(d,e)}^2/2 M_{\gamma}$ so that 
$\eta_{\gamma}^{(d,e)}$ are real. Then (${\cal M}, {\tilde {\cal R}}$ without channel 
indices refer to both `diagonal' and `nondiagonal' quantities):\\
1. The whole $\vartheta$-dependence of $\left|{\cal M}^{T^C}\right|$ is 
given by that of $\left|{\cal M}^{V^C} \right|$, 
\begin{eqnarray}
\left|{\cal M}^{T^C}(\vartheta, E) \right| \approx \left|{\tilde {\cal R}}^{(s)}(E) \right| 
\left|{\cal M}^{T^C}(\vartheta, E)\right|: \label{tir}
\end{eqnarray}
the CBA fails with respect to the magnitude only, insofar as $|{\tilde {\cal R}}^{(s)}(E) | $ 
differs from one.\\
2. With the penetration factor $C_{0}^2 = 2 \pi \eta_{\gamma}^{(d,e)}/[\mbox{exp}\{2 \pi
\eta_{\gamma}^{(d,e)}\}-1 ]$ pertaining to the intermediate-state Coulomb scattering one finds 
\begin{eqnarray}
\left|{\cal M}^{T^C} \right| \quad \stackrel{\eta_{\gamma}^{(d,e)} \to \,0}{=} 
\quad C_{0}^2 \left(1 + O({\eta_{\gamma}^{(d,e)}}^2)\right)
\left|{\cal M}^{V^C} \right| \quad \forall \, m,n,\vartheta. \label{bound} 
\end{eqnarray}
Since $C_{0}^2 < 1(> 1) $ for $e_{\alpha}e_{\beta} > 0(< 0)$, (\ref{bound}) not only quantifies the 
charge-sensitivity of $|{\cal M}^{T^C}| $ as compared to the insensitivity of 
$|{\cal M}^{V^C}| $ (i.e., ${\cal M}^{V^C}(e_{\alpha}e_{\beta} > 0) = -
{\cal M}^{V^C}(e_{\alpha}e_{\beta} < 0) $), but it also provides a simple method 
to quantitatively estimate the former. \\
3. Since $C_{0}^2(e_{\alpha}e_{\beta} > 0) = C_{0}^{-2} (e_{\alpha}e_{\beta} < 0) \left(1+O(
\eta_{\gamma}^{(s)2}) \right)$ one derives 
\begin{eqnarray} 
\left|{\cal M}^{T^C}_{\gamma,n m}(e_{\alpha}e_{\beta} > 0)\right| 
\left|{\cal M}^{T^C}_{\gamma,n m}(e_{\alpha}e_{\beta} < 0) \right| 
\approx \left({\cal M}^{V^C}_{\gamma,n m}\right)^2 \, \forall \, n, m, 
\vartheta, E, 
\end{eqnarray}
relating the `direct' triangle amplitudes for processes with opposite signs of the charges 
of the rescattering particles.\\
4. The dependence of $k_{(d,e)}$ on $n$ and $m$ becomes negligible for sufficiently large $E$, 
rendering $\eta_{\gamma}^{(d,e)}$ and $C_{0}^2$ practically independent 
of all bound state characteristics. Consequently, relation (\ref{bound}), and all results 
derived from it, become universal, i.e., state-independent. 

Many numerical tests of the approximate triangle amplitudes have been performed with
$e^{\pm},p,\bar p$ as projectiles and $H, P\!s, (\bar p,p)$ as targets \cite{akm96}. 
Quite generally one finds for both direct and exchange reactions that they reproduce 
the exact triangle amplitudes already 
at 1 keV incident energy for light, and at 100 keV for the heavy projectiles, 
to within a few percent, for practically all scattering angles, the agreement 
becoming even better with increasing energy. This is to be contrasted with the CBA's: 
not only are they real, but their magnitudes for light projectiles at 1 keV are off by 
$\sim$ 40 - 100 per cent, the improvement with increasing energy being slow only. 
Similarly for heavy projectiles.

\subsection{Proton-Deuteron Scattering with Realistic Potentials}

Ever since the development of the exact few-body theories, the investigation 
of the nucleon-deuteron ($Nd$) elastic scattering and breakup reaction 
has been at the center of interest, as their principal field of application. For 
processes with neutrons as projectiles (nd), highly sophisticated calculations 
are available nowadays, using realistic nuclear potentials including three-body forces. 
Apart from a few noticeable and as yet unexplained failures, they provide a good to very good 
description of all the available observables (for a recent review see \cite{gwhkg96}). 
However, experimental nd data are 
rather sparse and are lacking the desired accuracy. Hence, usually nd calculations 
are compared with pd data which makes too good an agreement rather questionable.

For proton-induced reactions (pd), on the other hand, data are abundant and of excellent 
precision, but similarly advanced calculations have hitherto been restricted to $E<0$ 
\cite{bsz88,krv94} (calculations with simple nuclear potentials, which have been performed 
for all energies, give in general "only" semiquantitative agreement but they are usually very 
accurate in explaining observed differences between pd and nd data, see \cite{asz85,ar94} and 
references therein). The reason for this lack of sophisticated pd calculations above 
the breakup threshold is obvious: In coordinate space, imposing Coulomb boundary 
conditions in the whole configuration space is a very difficult task. In momentum 
space, though the screening and renormalisation approach \cite{asz78,as80,as96} 
is in principle straightforward, its application is very computer time consuming 
(for a pedagogical introduction to the theory and a rather complete list of references, 
both on theory and calculations, see \cite{as96}).

We have now succeeded to obtain for the first time results for cross section and polarisation 
observables above the breakup threshold for a "realistic" (Paris) potential \cite{ams97} 
within the screening and renormalisation approach. In fact, we use a separable 
representation of the latter (so-called PEST1-6) known to provide an excellent 
approximation to the original local potential \cite{hp84}. Up to now the nucleon-nucleon 
interaction has been taken into account in the states $^3S_1-^3\!\!D_1$, 
$^1\!S_0$ and in all $P$ waves. When comparison is made with selected experimental 
data at 5 and 10 MeV \cite{sag94,sper84}, we find good although not perfect agreement 
of the calculated with measured quantities. Part of the discrepancies (in the polarisation 
observables $A_y$ and $i\,T_{11}$) exist already in the nd case, implying that there, as 
well as in our calculations, some aspects of the nuclear force are still missing. Other 
discrepancies may arise from the restriction in the number of 
nucleon-nucleon partial waves taken into account, or from a possible influence of the - hitherto 
neglected - three-body forces. However, the general trends, in particular concerning the 
relation between neutron and proton data, are well reproduced. Clearly, the observed agreement 
of nd-calculations with pd-data, in particular for $i\,T_{11}$ and also to a somewhat lesser extent 
for $A_y$, is purely fortuitous and does not arise from smallness of Coulomb 
effects in that observable.

A final remark concerns the fact that we have taken into account the Coulomb interaction 
in CBA, which for the present case has been estimated in \cite{akmr95,akm96} to 
be accurate to better than 1\%.\\[2mm] 

\noindent
{\bf Acknowledgment:} The research reported here has been financially supported by the 
Deutsche Forschungsgemeinschaft, Project 436 USB-113-1-0, by DOE Grant DE-FG05-93ER0773, 
and by Deutscher Akademischer Austauschdienst (DAAD).


\begin{thebibliography}{99}
\bibitem{ros72} L. Rosenberg, Phys. Rev. D {\bf 8} (1972) 1833.
\bibitem{am92} E. O. Alt, A. M. Mukhamedzhanov, JETP Lett. 
{\bf 56} (1992) 435; Phys. Rev. A {\bf 47} (1993) 2004.
\bibitem{ks96} Sh. D. Kunikeev, V. S. Senashenko, JETP {\bf 82} (1996) 839.
\bibitem{ml96} A. M. Mukhamedzhanov, M. Lieber, Phys. Rev. A {\bf 54} (1996) 3078.
\bibitem{km88} A. A. Kvitsinskii, S. P. Merkuriev, Sov. J. Nucl. Phys. {\bf 48} (1988) 79.
\bibitem{mss82} I. E. McCarthy, B. C. Saha, A. T. Stelbovics, Phys. Rev. A {\bf 22} (1980) 502.
\bibitem{uc89} K. Unnikrishnan, J. Callaway, Phys. Lett. {\bf A 138} (1989) 285.
\bibitem{am95} E. O. Alt, A. M. Mukhamedzhanov, Phys. Rev. A {\bf 51} (1995) 3852.
\bibitem{m97} A. M. Mukhamedzhanov, Phys. Rev. A {\bf 56} (1997) 473.
\bibitem{ves70} A. M. Veselova, Theor. Math. Phys. {\bf 3} (1970) 542.
\bibitem{ags67} E. O. Alt, P. Grassberger, W. Sandhas, Nucl. Phys. {\bf B2} (1967) 167.
\bibitem{maa97} A. M. Mukhamedzhanov, E. O. Alt, G. V. Avakov, Contribution 
to the International Conference on Few-Body Problems in Physics, Groningen, 1997; Preprint.
\bibitem{asz78} E.O. Alt, W. Sandhas, H. Ziegelmann, Phys. Rev. C {\bf 17} (1978) 1981. 
\bibitem{as80} E.O. Alt, W. Sandhas, Phys. Rev. C {\bf 18} (1980) 1088.
\bibitem{as96} E.O. Alt, W. Sandhas, Collision theory for two- and three-particle 
systems interacting via short-range and Coulomb forces: in Coulomb Interactions in 
Nuclear and Atomic Few-Body Collisions (F. S. Levin and D. Micha, eds.): Plenum, New York 1996.
\bibitem{akm96} E. O. Alt, A. S. Kadyrov, A. M. Mukhamedzhanov, Phys. Rev. A {\bf 54} (1996) 4091; 
Phys. Rev. A 53 (1996) 2438; J. Phys. B: At. Mol. Phys. 30 (1997) 3659.
\bibitem{akmr95} E. O. Alt, A. S. Kadyrov, A. M. Mukhamedzhanov, M. Rauh, J. Phys. B {\bf 28} (1995) 5137.
\bibitem{gwhkg96} W. Gl\"ockle, H. Witala, D. H\"uber, H. Kamada, J. Golak; Phys. Rep. {\bf 274} (1996) 107.
\bibitem{bsz88} G. H. Berthold, A. Stadler, H. Zankel, Phys. Rev. Lett. {\bf 61} (1988) 1077; 
Phys. Rev. C {\bf 41} (1990) 1365.
\bibitem{krv94} A. Kievsky, S. Rosati, M. Viviani, Nucl. Phys. {\bf A 577} (1994) 511; 
{\bf A607} (1996) 402.
\bibitem{asz85} E.O. Alt, W. Sandhas, H. Ziegelmann, Nucl. Phys. {\bf A 445} (1985) 429.
\bibitem{ar94} E.O. Alt, M. Rauh, Phys. Rev. C {\bf 49} (1994) R2285; 
Few-Body Syst. {\bf 17} (1994) 121.
\bibitem{ams97} E. O. Alt, A. M. Mukhamedzhanov, A. S. Sattarov, Post-deadline 
contribution to the International Conference on Few-Body Problems in Physics, Groningen, 1997.
\bibitem{hp84} J. Haidenbauer, W. Plessas, Phys. Rev. C {\bf 30} (1984) 1822.
\bibitem{sag94} K. Sagara et al, Phys. Rev. C {\bf 50} (1994) 576.
\bibitem{sper84} F. Sperisen F et al, Nucl. Phys. {\bf A 422} (1984) 81.
\end{thebibliography}
\end{document}